\documentclass[aps,prl,floats, twocolumn,showpacs,superscriptaddress,reprint]{revtex4-1}
\usepackage{graphicx,epsfig}
\usepackage[colorlinks=true,citecolor=blue]{hyperref}

\usepackage{graphics,dcolumn,bm,epic, eepic,fleqn,float}
\usepackage{amssymb,amsmath,amsfonts,multirow,rotate,color}
\usepackage{soul}
\usepackage{graphicx}
\definecolor{Myorange}{cmyk}{0,0.42,1,0}

\begin{document}
\title{Stabilizing synchrony by inhomogeneity}

\author{Ehsan Bolhasani}
\affiliation{Department of physics, Institute for Advanced Studies in Basic Sciences, Zanjan, Iran.}
\affiliation{School of Cognitive Sciences, Institute for Studies in Theoretical Physics and Mathematics, Niavaran, Tehran, Iran.}
\author{Alireza Valizadeh}
\affiliation{Department of physics, Institute for Advanced Studies in Basic Sciences, Zanjan, Iran.}
\affiliation{School of Cognitive Sciences, Institute for Studies in Theoretical Physics and Mathematics, Niavaran, Tehran, Iran.}

\date{\today}

\begin{abstract}
 We show that for two identical neuronal oscillators with strictly postive phase restting curve, isochronous synchrony is an unstable attractor and arbitrarily weak noise can destroy entraiment and generate intemittent phase slips. Small inhomogeneity--mismatch in the intrinsic firing rate of the neurons-- can stabilize the phase locking and lead to more precise relative spike timing of the two neurons. The results can explain how for a class of neuronal models, including leaky itegrate-fire model, inhomogeneity can increase correlation of spike trains when the neurons are synaptically connected.
\end{abstract}

\pacs{05.45.Xt, 87.19.lm, 89.75.Fb, 89.75.Kd}

\maketitle

Synchronization observed frequently in the vast variety of physical, chemical, industrial, and biological complex systems, from coupled pendulum clocks to neuronal populations in the nervous system \cite{huygens1966horologium, bennett2002huygens, kuramoto2003chemical, buck1968mechanism, niedermayer2008synchronization, rohden2012self, mahboob2008bit, antonio2012frequency, sharma2012amplitude, gray1989oscillatory}. In these systems, ability to exhibit synchronous or phase locked oscillations, is the foundation of the emergent behaviors which is the basis for the functionality of the system. While the existence of robust synchronization is important in real systems with different sources of noise and uncertainties in parameters, stability of this behavior has a central importance \cite{buck1968mechanism}. 
Recordings of multi-neuron spike trains have revealed significant interdependencies between the firing of different neurons in a population \cite{zohary1994correlated, meister1995concerted, alonso1996precisely, christopher1996primary, bair2001correlated, kohn2005stimulus}. Synchronous oscillations are found in many brain regions and excessive synchrony is a hallmark of neurological disorders such as epilepsy and Parkinson's disease \cite{pyragas2007controlling}. Functional role of the correlation in neural coding has been debated in recent years \cite{abbott1999effect, nirenberg2001retinal, nirenberg2003decoding, averbeck2006neural, schneidman2006weak, pillow2008spatio}. Synchrony itself may encode information directly \cite{christopher1996primary, kohn2005stimulus, de2007correlation, gray1989oscillatory, biederlack2006brightness, chacron2008population, josic2009stimulus}. Synchronous firing of the neurons in one region serves to reliably transmit signals to upstream regions \cite{salinas2000impact, kuhn2003higher, tetzlaff2008dependence}, while synchrony between different regions can prepare dynamic channels for communication \cite{fries2005mechanism, fries2007gamma, fries2009neuronal} and also undelies feature binding \cite{singer1999neuronal}. Beyond the functional role, it is also important to understand how correlation and synchrony depend on biophysical parameters of the neurons and the network. 

Correlation between spike trains of neurons can arise from shared input they receive from other neurons \cite{sears1976short, binder2001relationship, constantinidis2001coding, turker2001effects, turker2002effects}, or from presence of direct synaptic connections between neurons \cite{snider1998burst, csicsvari1998reliability, bartho2004characterization, fujisawa2008behavior}. In both cases the collective state of the system depends on the parameter of neurons, e.g., firing rate and the type of the excitability of neurons \cite{abouzeid2009type}, and the parameters of the connections such as delay \cite{sadeghi2014synchronization}. Physiological heterogeneity can destabilize both coupling-induced and correlation-induced synchronization \cite{kopell2002mechanisms, burton2012intrinsic, santanello2013impact}. In the classical models of synchronization, collective state of a system of coupled oscillators is determined by outcome of rivalry between synchronizing effect of connections and desynchronizing effect of inhomogeneity \cite{kuramoto2003chemical}, but there are examples of the systems in which synchrony is enhanced by inhomogeneity \cite{braiman1995disorder, valizadeh2007fractional}. Recently we have shown that small inhomogeneity can increase correlation between spike trains of two coupled neurons \cite{bolhasani2013direct}. In this sudy we give a general framework for the correlation of coupled phase oscillators with a given phase sensitivity. We show that for identical pulse coupled type-I oscillators, synchronized state is an unstable attractor and arbitrarily weak noise can destabilize this state and the spiking of two neurons exhibit intermittent phase slips between epochs of locking. Small inhomogeneity in firing rates can stabilize the system by providing an asymmetric basin of attraction around the stable phase-locked state. This in turn results in a sharper PDF for the time difference between spikes of the two neurons in presence of noise. We have also shown that while for the model neurons with biologically realistic phase response curve (PRC), the time difference between the spikes of two neurons in the stable state increases with inhomogeneity, in the case of LIF neurons, they lock in almost zero phase lag for sufficiently small values of inhomogeneity. By solving Fokker-Planck equation we also find the most probable phase difference between spike times of the two neurons and will show that it does not coincide with the stable point of the deterministic equations. 
 
Our model comprises two bidirectionally coupld neuronal oscillators recieving suprathreshold constant currents ($I_1$ and $I_2$ with mismatch $\Delta I$) as well as independent stochastic inputs. The evolution of the state vector of the oscillators $X_i, i=1,2$ can be descibed by
\begin{eqnarray}
 \dot{X}_1  = F(X_1)+\epsilon g_{12}G_{12}(X_1,X_2)+I_1+\sigma \xi_1(t)\\
\dot{X}_2  = F(X_2)+\epsilon g_{21}G_{21}(X_2,X_1)+I_2+\sigma \xi_2(t),
\label{model}
\end{eqnarray} 
where $F$ governs the internal dynamics of the neurons, $G$ detemines the synaptic connections, $\xi$ is Gaussian white noise with zero mean and unit variance, and $\epsilon$ and $\sigma$ are small values which scale strength of the couplings and the stochatic inputs, respectively. We assume the each of the unperturbed systems $ \dot{X}_i  = F(X_i)$ has an asymptotically stable limit cycle, $X_0(t) = X_0 (t + T)$, so that a phase variable can be defined in vicitnity of the limit cycle. In the regime of weak coupling and weak noise we can apply the standard phase reduction \cite{kuramotochemical, ermentrout1996type, teramae2004robustness} to the Langevin equations above. The system is then can be described by a set of $\rm{It{\hat{o}}}$ stochastic differential equations:
\begin{eqnarray}
\dot{\theta}_1  = \omega_1 &+&\varepsilon  g_{12}Z\left(\theta _{1}\right) G\left(\theta _{1}, \theta _{2} \right) \nonumber \\ &+& \sqrt{D \varepsilon } Z\left( \theta _{1}\right) \xi _1 \left( t \right)  \nonumber \\
\dot{\theta}_2  = \omega_2 &+&\varepsilon  g_{21}Z\left(\theta _{2}\right) G\left(\theta _{2}, \theta _{1} \right) \nonumber \\  &+& \sqrt{D \varepsilon } Z\left( \theta _{2}\right) \xi _2 \left( t \right) 
\label{ophase}
\end{eqnarray} 
where $Z(\theta)$ is the infinitesimal phase-response curve (PRC) \cite{winfree2001geometry}. We assume that the natural frequencies have a small difference $\omega _{1} -\omega_2= \Delta \omega$ and the noise and coupling influence only the first (voltage) variable of the state vector of the neural oscillators. In our model the neurons communicate via pulsatile signals $G_{ij}=\sum_n \delta(t-t^n_j)$, where $\delta$ is Dirac's delta function and $t^n_j$ is the instant of $n^{th}$ firing of the neuron $j$. These pulses idealize the communcation signals which are short compared to the intrinsic time scale of the oscillators and are used to model diverse systems such as populations of flashing fireflies and plate tectonics in earthquakes, as well as networks of spiking neurons in the brain \cite{mirollo1990synchronization, gerstner1996rapid, hansel2001existence, timme2002prevalence, peskin1975mathematical, strogatz2001exploring}. It is assumed that the mismatch, coupling and noise terms are of the same order, sufficiently weak such that the intrinsic dynamics of isolated identical phase oscillators is dominant. 

\begin{figure}[h]
\centering
\includegraphics[width=3.4in,angle=-0]{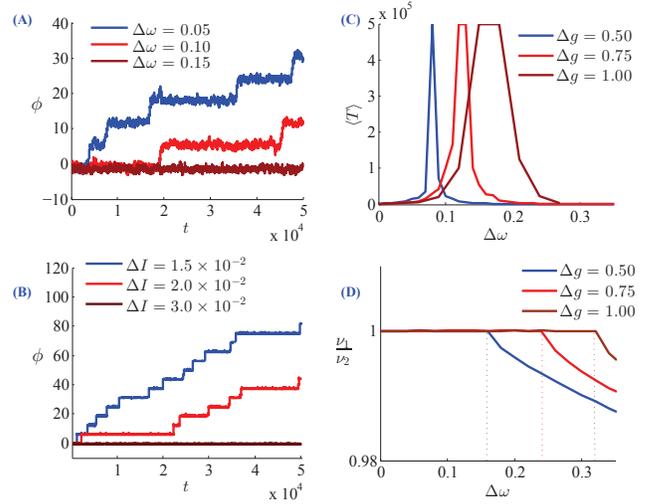}
\caption{(A,B) Representative examples of the evolution of the phase difference of two neurons for three different values of mismatch in intrinsic frequencies. Larger values of mismatch have led to fewer phase slips. In A neurons are phase oscillators with canonical type-I phase sensitivity and in B the results are presented for LIF neurons. (C) The mean scape time is plotted against frequency mismatch. Increasing effective coupling constant $\Delta g= g_1-g_2$ the maximum scape time is seen in largar values of frequency mismatch. In (D) the ratio of the firing rates of the coupled neurons is plotted. For large values of mismatch the fixed point of $1:1$ locking vanishes. }
\label{fig1}
\end{figure}

Using the method of averaging \cite{ermentrout1991multiple} we derive the the equation of motion for the phase difference $\phi= \theta_1- \theta_2$:
\begin{eqnarray}
\dfrac{d\phi}{dt} & = & \varepsilon \left[ \Delta\omega -g_{12}H\left( \phi \right) + g_{21}H\left(- \phi \right) \right] \nonumber \\ & + &\sigma _{\phi} \sqrt{2D\varepsilon}~ \eta \left( t \right)
\label{phasediff}
\end{eqnarray} 
where $ H \left(  \phi  \right)  =  \frac{1}{T}\int _{0}^{T} Z \left(\tilde {t} \right) G\left( \tilde {t}, \tilde {t}+\phi \right) d\tilde{t}$ and the coefficient $\sigma _{\phi}= \left( \frac{1}{T} \int _{0}^{T} \left[ Z \left( \tilde{t} \right) \right] ^{2} d\tilde{t} \right) ^{1/2}$ comes from averaging the noisy phase equations \cite{kuramotochemical}. Here $\eta \left( t \right) = \frac{\xi _{1} \left( t \right) - \xi _{2} \left( t \right)}{\sqrt{2}}$ is itself a Gausian white noise with zero mean and unit variance.

We restrict the study to type-I oscillators and first discuss on the deterministic version of Eq. \ref{phasediff} with $D=0$. If $H$ is an even function of $\phi$, e.g. for QIF oscillators, the effective coupling term would be $\Delta g H(\phi)$ with $\Delta g= g_{21}-g_{12}$. In this case the most effective connection is a unidirectional one and the symmetric connection has no effect on the relative dynamics of the oscillators. Note that for the oscillators with an {\it oblique} PRC, e.g. the LIF oscillators, the coupling term can be non-zero for symmetric connections (see suplementary material Fig.~S1). 

The fixed point of Eq. \ref{phasediff} with $D=0$ is the solution of $\Delta\omega =g_{12}H\left( \phi \right) - g_{21}H\left(- \phi \right)$. For QIF oscillators $Z \left( \phi \right) = 1-cos \left( \phi \right)$ with asymmetric connections $\Delta g \neq 0$, when the oscillators are identical $\Delta\omega =0$, the zero-lag synchrony $\phi=0$ is an unstable attractor and in the absence of noise, the oscillators can synchronize isochronously. But a waek noise can destroy synchrony and lead to phase slips. Mismatch in the intrinsic firing rates of the neurons, stabilizes the fixed point through a saddle-node bifurcation while moves the fixed point away from zero. For small mismatch, this provides an asymmatric basin of attraction which is vulnerable to sufficiently large perturbations in one direction around the fixed point. In the presence of noise the system shows epochs of intermittent locking between which the relative phase of the oscillators slips by one cycle, while the mean scape time from locked states increases with frequency mismatch (see Figs.~1A and B). The maximum mean scape time from the locked state, is seen in a certain value of mismatch (Fig.~ 1C) and for larger mismatches $\Delta\omega > Max \{g_{12}H\left( \phi \right) - g_{21}H\left(- \phi \right)\}$, the fixed point corresponding to $1:1$ locked state will disappear through another saddle-node bifuracation (Fig.~3). 
To give more concrete results on the impact of the inhomogeneity on the correlation of the spike trains of the neuronal oscillators in presence of noise, we derive the Fokker-Planck equation for the distribution of the phase difference of two neurons, described by Eq. \ref{phasediff}. We rewrite Eq. \ref{phasediff} in a more closed form
\begin{eqnarray}
\dfrac{d\phi}{dt} =\varepsilon \Delta g \Gamma \left( \phi \right)  + \sigma _{\phi} \sqrt{2D\varepsilon} \eta \left( t \right)
\label{phdiff_I}
\end{eqnarray}
 where $\Gamma \left( \phi \right) = \left[ \frac{\Delta\omega}{\Delta g} -\frac{1}{T} + \frac{1}{T} \cos \phi \right] $. The corresponding Fokker-Planck equation takes the form:
\begin{eqnarray}
\frac{\partial\rho}{\partial t} \left( \phi , t \right) = &-&\varepsilon\frac{\partial}{\partial \phi} \left[ \Gamma \left( \phi \right) \rho \left( \phi , t \right) \right] \nonumber \\ & + & \left( \sigma _{\phi}^{2}D\varepsilon  \right)\frac{\partial ^{2}\rho}{\partial \phi ^{2}}\left( \phi , t \right)
\label{Fokker}
\end{eqnarray}
where $\rho \left( \phi , t \right)$ is the distribution of the phase differences. Stationary solution of this equation with periodic boundary condition is:   
\begin{eqnarray}
\rho \left( \phi \right) &=& \dfrac{1}{N} e^{M\left( \phi \right)}  \nonumber \\ & \times &  \left[ \frac{e^{-\frac{1}{\alpha} \left( \frac{T \Delta\omega}{\Delta g}  -1 \right)} - 1}{\int _{0}^{T} e^{-M\left( \tilde{\phi} \right)} d\tilde{\phi}} \int _{0}^{\phi } e^{-M \left( \tilde{\phi} \right) } d\tilde{\phi} + 1\right]
\label{Ststate}
\end{eqnarray}
where $M\left( \phi \right) = \frac{1}{\alpha} \int _{0}^{\phi} \Gamma \left( \tilde{\phi} \right) d\tilde{\phi}$. Also, $N$ is the normalization factor so that $\int_{0}^{T} \rho \left( \phi \right) d\phi = 1$ and $\alpha =\frac{D\sigma _{\phi}^{2}}{\Delta g}$ is the ratio of noise intensity to the coupling strength \cite{netoff2012experimentally}.
\begin{figure}[h]
\centering
\includegraphics[width=3.6in,angle=-0]{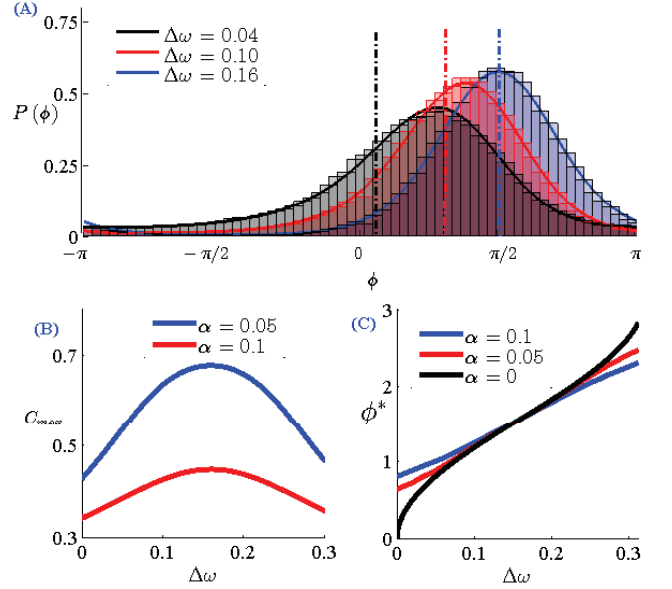}
\caption{(A) The steady state phase difference distributions $\rho \left( \phi \right)$ for three levels of heterogeneity. Distributions have became narrower as mismatch is increased. Solid lines show the analytic result Eq. \ref{Ststate} and the bar graph presents the numerical results by direct integration of Eqs. \ref{ophase}. Dashed vertical lines show the position of the fixed points of deterministic equations. (B) The maximum value of $\rho \left( \phi \right)$ is plotted against frequency mismatch for two different values of the the ratio of noise strength to effective coupling $\alpha =\frac{D\sigma _{\phi}^{2}}{\Delta g}$. (C) The most probable phase difference are shown for two values of $\alpha$. They don't concide with the location of fixed point of the deterministic equations (black curve). }
\label{fig2}
\end{figure}
\begin{figure}[h]
\centering
\includegraphics[width=3.6in,angle=-0]{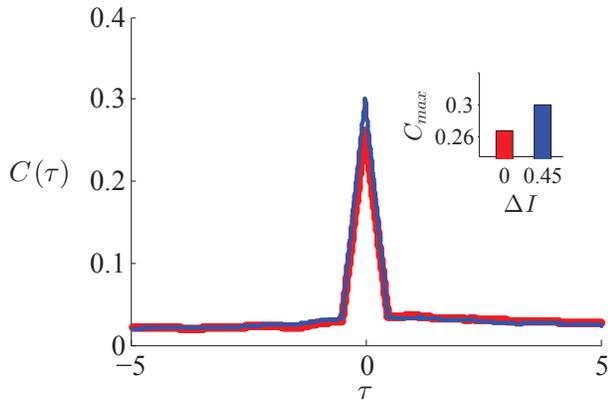}
\caption{The cross-correlogram of spike trains of two LIF neurons $C \left(\tau \right)$ shows that in presence of the mismatch cross correlation is increased. The level of maximum correlation is shown in the inset to highlight the increase due to the inhomogeneity.}
\label{fig3}
\end{figure}

Figure 2A shows the steady state phase difference distribution for different values of the frequency mismatch for QIF neuronal oscillators. It can be seen that the distribution becomes narrower (with a more pronounced peak) with increasing frequency mismatch while the neurons remain in $1:1$ locked state, i.e. for the mismatch in the range $0\leq \frac{\Delta \omega}{\Delta g} \leq \frac{2}{T}$. This reflects a larger basin of attraction for the locked state when mismatch is increased from zero. Furthermore, the asymmetry of the basin of attraction causes the distribution of the phase differences not to peak in the fixed point of the deterministic equation, determined by $\Gamma \left( \phi ^{*} \right) = 0$. In turn, in presence of noise the location of maximum phase difference satisfies,
\begin{eqnarray}
\Gamma \left( \phi ^{*} \right) =\alpha \frac{1-e^{-\frac{1}{\alpha} \left( \frac{T \Delta \omega}{\Delta g}-1 \right)}}{\rho \left( \phi ^{*} \right) \int _{0}^{T} e^{-M\left( \bar{\phi} \right)} d\bar{\phi}}.
\label{gamaphi}
\end{eqnarray}
which is derived by taking the derivative of $\rho \left( \phi \right)$ with respect to $\phi$, equal to zero. The location of the most probable phase difference as a function of mismatch, detrermined by Eq. \ref{gamaphi}, is plotted in Fig. 2C for different values of noise amplitude as well as for noiseless system which shows the location of the fixed point. Presence of noise inclines the distribution to larger phase differences for small values of frequency mismatch. The maximum difference between the location of most probable phase difference between noiseless state and noisy state is seen near $\frac{\Delta \omega}{\Delta g} = 0 \rm{~or~} \frac{2}{T} $ which reflects the most asymmetric basin of attraction for the locked state and in turn the locations coincide when $\phi^*=\pi/2$ where the basin of attraction is symmetric.

Pfeuty {\it et al.} (2005) have introduced a variable $S_{i} \left( t \right)$ which is equal to $1/ \delta$ when a neuron has fired a spike in a time bin of size $\delta $ about time t and is equal to $0$ otherwise \cite{pfeuty2005combined}. For sufficiently small $\delta$ the time average of $S_{i}$ is the average firing rate of neuron $i$. It is shown that the normalized cross-corellogram (CC) of this variable which is the density of probability that neuron 2 to fire a spike in a time bin of size $\delta$ a delay $\tau$ after a spike of neuron 1, is related to the phase difference probability distribution function $\rho \left( \phi \right)$ through
\begin{eqnarray}
\rho \left( \frac{\tau}{T} \right) = C \left( \tau \right) = \frac{\langle S _{1} \left( t \right) S_{2} \left( t+\tau \right) \rangle}{\langle S_{1} \left( t \right) \rangle \langle S_{2} \left( t \right) \rangle}
\label{Pfeuty}
\end{eqnarray}
where $\langle ... \rangle$ indicates averaging over time. A peak in CC at a time lag $\tau$ shows phase locking of the activity of the neurons. The sharper CC is indicator of a tighter locking. To illustrate this effect we provide an expression for the maximum value of the distribution fuction (or CC) as a function of frequency mismatch:
\begin{eqnarray}
C_{Max}\left( \frac{\Delta \omega}{\Delta g} \right) =\alpha \frac{1-e^{-\frac{1}{\alpha} \left(  \frac{T\Delta \omega}{\Delta g}-1\right)} } { \Gamma \left( \phi ^{*} _{\left( \frac{\Delta\omega}{\Delta g}\right)} \right) \int _{0}^{T} e^{M\left( \bar{\phi} \right)}d \bar{\phi}} 
\label{maxp}
\end{eqnarray}
The above equation is the same as Eq. \ref{gamaphi} by substituting $\rho \left( \phi ^{*} \right)$ with $C\left( \frac{\Delta \omega}{\Delta g}\right)$. Figure 2B shows maximum value of cross-correlation versus frequency mismatch for different values of noise to coupling ratio which is resulted from direct integration of Eq. \ref{gamaphi}. The result shows that the maximum cross-correlation of the spike trains of the oscillators would be also maximum when the neurons are not identical. This is a consequence of more precise relative spike timing of the two neurons in presence of inhomogeneity. 

In this study we have shown that for two synaptically connected neuronal oscillators, more precise relative spike timing can be achieved when the neurons receive different levels of inputs and have different intrinsic firing rates. Consequently, cross-correlation of spike trains of the neurons increases in presence of mismatch in intrinsic firing rates of neurons. While the results are presented for neuronal oscillators, they can find application in general context of coupled limit cycle oscillators.

\bibliographystyle{apsrev}
\bibliography{Mypaper}

\pagebreak
\widetext
\begin{center}
\textbf{\large Supplementary Material for\\ ``Stabilizing synchrony with heterogeneity''}
\end{center}
\setcounter{equation}{0}
\setcounter{figure}{0}
\setcounter{table}{0}
\setcounter{page}{1}
\makeatletter
\renewcommand{\theequation}{S\arabic{equation}}
\renewcommand{\thefigure}{S\arabic{figure}}
\renewcommand{\bibnumfmt}[1]{[S#1]}
\renewcommand{\citenumfont}[1]{S#1}

\section{Weakly coupled oscillators}
Our model comprises two bidirectional coupled neurons receiving suprathreshold constant currents as well as uncorrelated stochastic inputs.  The general form of equations describing this model is given by
\begin{eqnarray}
\dot{X}_1  = F(X_1)+\epsilon g_{12} G_{12}(X_1,X_2)+I_1+\sigma \xi_1(t) \nonumber \\
\dot{X}_2  = F(X_2)+\epsilon g_{21} G_{21}(X_2,X_1)+I_2+\sigma \xi_2(t)
\label{eqx}
\end{eqnarray} 
where $X_{i}$ is a N-dimensional state vector containing the membrane potential and gating variables. For example in the multicompartmental Hodgkin-Huxley (HH) model, $X = \left[ V,m,h,n \right] ^{T}$ and in the single compartmental Leaky Integrate-and-Fire (LIF) model, $X = V$. $F\left( X \right)$ defines the internal dynamics of the neuron $i$. $G_{ij}\left( X_{i},X_{j} \right)$ determins functional form of synaptic connection from neuron $j$ to neuron $i$. For example  in the case of pulse coupled synapses, as we used in this letter it would be 
\begin{eqnarray}
G_{ij}\Big( X_{i}\left( t \right),X_{j}\left( t \right) \Big) = \sum _{n} \delta \left( t - t^{n}_{j} \right)
\label{synapseeqx}
\end{eqnarray}
where $t_{j}^{n}$ is the instant of $n^{th}$ spike of neuron j and $\delta \left(t\right)$ is the Dirac's delta function. $g_{ij}$ shows the synaptic weight from neuron $j$ to neuron $i$ which is scaled by a small factor $\varepsilon$ so that the weakly coupled oscillators approximation is valid.


Each neuron receives suprathreshold constant current $I_{i}$ with mismatch $\Delta I = I_{1} - I_{2}$ as well as an independent Gaussian white noise characterized by its mean, auto- and cross-correlation with the other neuron's input
\begin{eqnarray}
\langle \xi _{i} \left( t \right) \rangle = 0,
\end{eqnarray}
\begin{eqnarray}
\langle \xi _{i} \left( t \right) \xi _{j} \left( t' \right) \rangle = \delta _{ij} \delta \left( t-t' \right).
\end{eqnarray}
We assume that at the absence of synaptic connections and noise, dynamics of isolated neurons given by
\begin{eqnarray}
\dot{X}_i  = F(X_i) + I_i
\end{eqnarray}
have a T-periodic limit cycle solution $X_{0} \left( t\right) $. We assumed that the magnitude of the coupling term scaled with a small coefficient $\varepsilon$ and also the amplitude of the noise is such that the variance of the noise and the strength of the coupling is in the same order, so $\sigma = \sqrt{D\varepsilon }$, where $D= \mathcal{O}\left( 1 \right)$. In such a weak coupling and weak noise regime the dynamics of the neurons can be approximated by defining a single phase variable around limit cycle. So we define a phase variable, $\theta _{i} \left( t \right) $ in the vicinity of unperturbed limit cycle for each oscillator and reduce high dimensional Eqs. \ref{eqx} to two scalar equations for the evolution of the phase 

\begin{eqnarray}
\dot{\theta}_1  = \omega_1 + \varepsilon g_{12}Z\left(\theta _{1}\right) G\left(\theta _{1}, \theta _{2} \right)  + \sqrt{D \varepsilon } Z\left( \theta _{1}\right) \xi _1 \left( t \right)\\
\dot{\theta}_2  = \omega_2 + \varepsilon g_{21}Z\left(\theta _{2}\right) G\left(\theta _{2}, \theta _{1} \right)  + \sqrt{D \varepsilon } Z\left( \theta _{2}\right) \xi _2 \left( t \right)  
\label{ophase}
\end{eqnarray} 

where $Z\left( \theta \right)$ is the infinitesimal phase response curve (PRC). We assumed that neurons natural frequencies have small difference, $\varepsilon\Delta\omega = \omega _{2} - \omega _{1}$, where $\varepsilon$ confirms that the mismatch is of the order $\mathcal{O}\left( \varepsilon \right)$. Assuming $X_{i} \left( t \right) \simeq X_{0} \left( \theta _{i} \left( t \right) \right)$ and by the change of variable $\theta _{i} \left( t \right) = \omega t+ \phi _{i} \left( t \right)$, the equations for the evolution of the relative phase of the oscillators read:
\begin{eqnarray}
\dot{\phi_ {1}} &=& -\dfrac{\varepsilon \Delta \omega}{2} + \varepsilon g_{12}Z\left( \phi _{1} \right) G\left( \phi _{1},\phi _{2} \right) + \sqrt{D\varepsilon} Z\left( \phi _{1} \right) \xi _{1} \left( t \right)\\
\dot{\phi_ {2}} &=& +\dfrac{\varepsilon \Delta \omega}{2} + \varepsilon g_{21}Z\left( \phi _{2} \right) G\left( \phi _{2},\phi _{1} \right) + \sqrt{D\varepsilon} Z\left( \phi _{2} \right) \xi _{2} \left( t \right).
\label{eqnphase}
\end{eqnarray}

We exploit the fact that $\varepsilon$ is small  to further reduce Eqs. \ref{eqnphase}. With a system of the form
\begin{eqnarray}
\dot{x} = \varepsilon f \left( x, t \right).
\label{ave}
\end{eqnarray}
Averaging theory states that in Eq. \ref{ave}, $x\left( t \right)$ can be replaced by its average over a period $\bar{x}$ and 
\begin{eqnarray}
\dot{\bar{x}}  = \varepsilon \dfrac{1}{T}\int _{0}^{T} f\left( \bar{x}, t\right) dt.
\end{eqnarray}
By applying averaging method on the Eqs. \ref{eqnphase} we have
\begin{eqnarray}
\dot{\phi _{1}} &=& - \varepsilon \dfrac{\Delta \omega}{2} + \varepsilon \dfrac{g12}{T} Z \left( \phi _2 - \phi _1 \right)  + \sigma _{\phi}\sqrt{D\varepsilon }  \xi _1 \left( t \right)\\
\dot{\phi _{2}} &=& +\varepsilon \dfrac{\Delta \omega}{2} + \varepsilon \dfrac{g21}{T} Z \left( \phi _1 - \phi _2 \right)  + \sigma _{\phi}\sqrt{D\varepsilon }  \xi _2 \left( t \right)
\end{eqnarray}

where the term $\sigma _{\phi} = \Big(  \frac{1}{T}\int _{0}^{T}  \left[ Z\left( \tilde{t} \right) \right] ^{2} d \tilde{t} \Big) ^{1/2}$
originates from averaging the noisy phase equations, and $Z \left( \phi _i - \phi _j \right)$ comes from
\begin{eqnarray}
H \left( \phi _2 - \phi _1 \right) &=& \dfrac{1}{T} \int _{0}^{T}   Z\left( \theta _{1} \right) G\left( \theta _{1}, \theta _{2} \right) d\theta _{1}\\
 &=& \dfrac{1}{T} \int _{0}^{T}   Z\left( \theta _{1} \right) \delta \left( \theta _{2} \right) d \theta _{1} 
\\&=& \dfrac{1}{T} \int _{0}^{T} Z\left( t+\phi _{1} \right) \delta \left( t+\phi _{2} \right) d t  
\\&=& \dfrac{1}{T} \int _{0}^{T}  Z\left( \tilde{t} \right)  \delta \left( \tilde{t} -\phi _{1} + \phi _{2} \right) d \tilde{t} 
\\&=&  \dfrac{Z\left( \phi _{2} - \phi _{1} \right)}{T}.
\end{eqnarray}
without loss of generality  we assumed the phase is normalized so that $0\leq\theta < T$, i.e., $\omega = 1$. By defining $\phi = \phi _2 - \phi _1$, we derive the following equation for the phase difference
\begin{eqnarray}
\dot{\phi} = \varepsilon \Big( \Delta \omega + g_{21}H(-\phi) - g_{12}H(\phi) \Big) + \sigma _{\phi} \sqrt{2D\varepsilon }  \eta \left( t \right) 
\label{phasedifference}
\end{eqnarray}
where $\sqrt{2}\eta \left( t \right) = \xi _{2} - \xi _{1}$ and $\eta (t)$ is a Gaussian white noise with zero mean and unit variance.

\begin{figure}[h]
\centering
\includegraphics[width=4.5in,angle=-0]{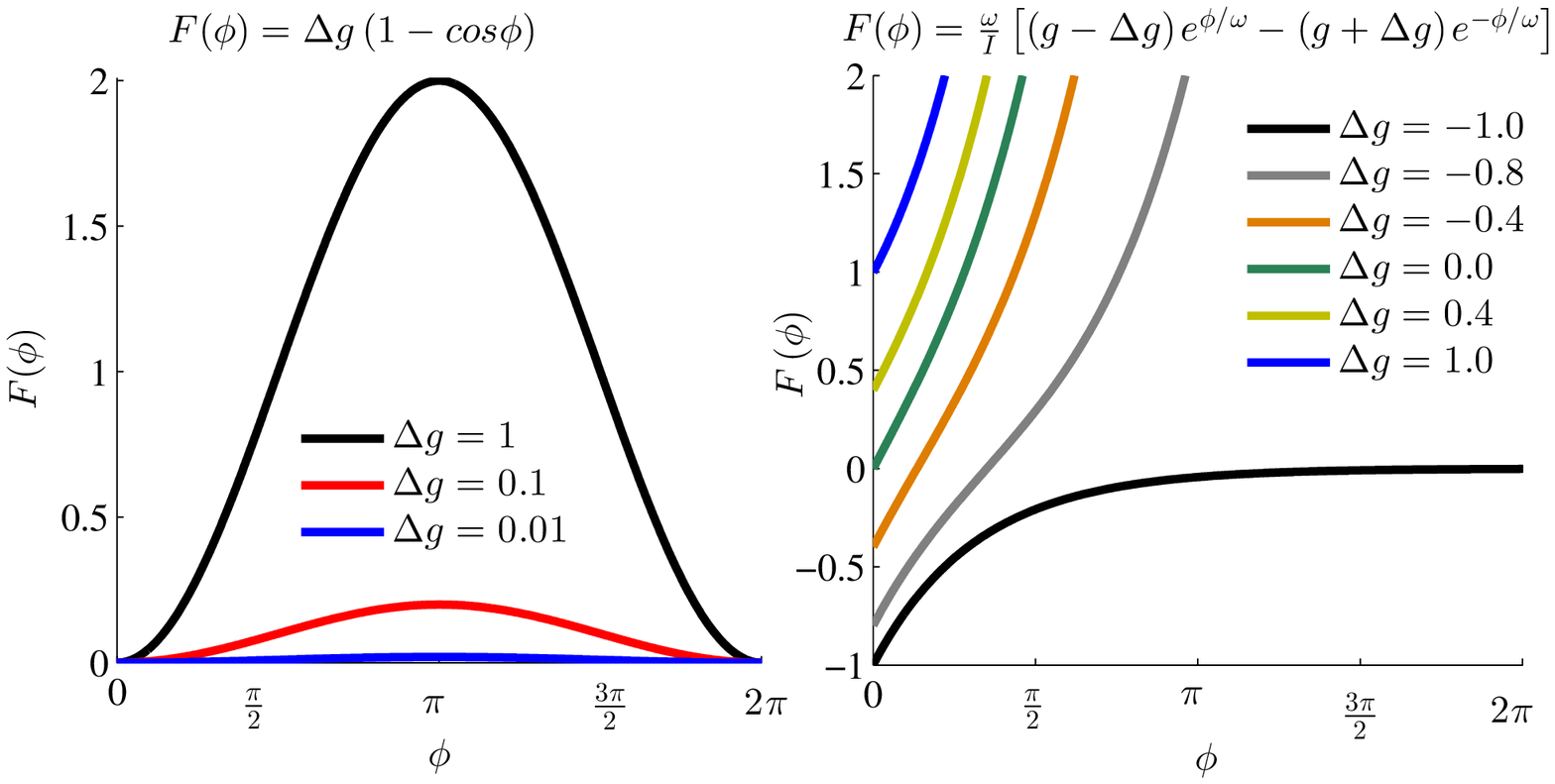}
\caption{Examples of $T\times F$ function, $F\left(\phi\right)=g_{12}H\left( \phi \right) - g_{21}H\left( -\phi \right)$, for QIF oscillator (left) and LIF oscillator (right) for different values of difference of the synaptic strengths. For the LIF oscillators with an uneven PRC the coupling term can be non-zero for symmetric connections whereas for QIF oscillators with an even PRC the effective coupling is determined by the difference of two reciprocal synaptic constants.}
\label{fig1}
\end{figure}

The have used two model neurons in ot study: Canoncal type-I oscillators with y $Z (\phi) = 1 - \cos (\phi)$ and LIF oscillators which is described by $\dot{v} = I - v$ for $t\leq v_{Th}(=1)$ and $lim _{\tau \rightarrow 0^{+}} v(t+\tau) = 0$ with
\begin{eqnarray}
Z (\phi) = \dfrac{\omega}{I} exp\Big( \dfrac{\phi}{\omega} \Big), ~~~0\leq \phi < 2\pi,
\end{eqnarray}     
where $\omega = 2\pi / [ logI - log(I - 1) ]$.
First we focus on the deterministic case of Eq. \ref{phasedifference} with $D=0$. For QIF oscillator, $Z \left( \phi \right) = 1- \cos \left( \phi \right)$ is an even function of $\phi$. Therefore it reduces to 
\begin{eqnarray}
\dot{\phi} = \varepsilon \left[ \Delta \omega + \dfrac{\Delta g}{T}\Big( 1- \cos (\phi ) \Big) \right] 
\label{phasedifferenceQIF}
\end{eqnarray}
where $\Delta g = g_{21} - g_{12}$ is the effective coupling constant.  In this case the most effective coupling is that which maximizes $\Delta g$. i.e., a unidirectional one and the symmetric connection leads to zero effective coupling $\Delta g=0$. But for LIF neurons with uneven PRC Eq. \ref{phasedifference} takes the form
\begin{eqnarray}
\dot{\phi} = \varepsilon \left[ \Delta \omega + \dfrac{\omega g_{21}}{IT} exp\Big( \dfrac{-\phi}{\omega} \Big) - \dfrac{\omega g_{12}}{IT} exp\Big( \dfrac{\phi}{\omega} \Big) \right] ~~~0\leq \phi < 2\pi.
\label{phasedifferenceLIF}
\end{eqnarray}
 Note that for the oscillators with an {\it oblique} PRC, e.g. the LIF oscillators, the effective coupling term can be non-zero for symmetric connections (see Fig.~\ref{fig1}). 
The fix points of general equation \ref{phasedifference} with $D=0$ are the cross points of horizontal line $T\Delta \omega$ and the curve described by $F(\phi) = g_{12} Z(\phi) - g_{21} Z(- \phi)$. 

For type-I phase oscillators we rewrite Langevin Eq. \ref{phasedifference} as
\begin{eqnarray}
\dfrac{d\phi}{dt} =\varepsilon Z g \Gamma \left( \phi \right)  + \sigma _{\phi} \sqrt{2D\varepsilon} \eta \left( t \right)
\label{phdiff_I}
\end{eqnarray}  
with $\Gamma \left( \phi \right) = \left[ \frac{\Delta\omega}{\Delta g} -\frac{1}{T} + \frac{1}{T} \cos \phi \right] $. Corresponding Fokker-Planck equation for the phase difference distribution $\rho (\phi ,t)$ is
\begin{eqnarray}
\frac{\partial\rho}{\partial t} \left( \phi , t \right) = -\varepsilon\frac{\partial}{\partial \phi} \left[ \Gamma \left( \phi \right) \rho \left( \phi , t \right) \right]  +  \left( \sigma _{\phi}^{2}D\varepsilon  \right)\frac{\partial ^{2}\rho}{\partial \phi ^{2}}\left( \phi , t \right).
\label{Fokker}
\end{eqnarray}   
The stationary phase difference distribution satisfies 
\begin{eqnarray}
\frac{\partial\rho _{0} \left( \phi , t \right)}{\partial t}  =0,
\end{eqnarray}
with the solution
\begin{eqnarray}
\rho _{0} \left( \phi \right) = \dfrac{1}{N}e^{M(\phi)} \left[ A \int _{0} ^{T}  e^{-M(\phi ')} d \phi ' + 1 \right],
\end{eqnarray}
where 
\begin{eqnarray}
M(\phi) =\dfrac{1}{\alpha } \int _{0} ^{\phi} \Gamma \left( \bar{\phi} \right) d \bar{\phi}.
\end{eqnarray}
$N$ is a normalization factor so that $\int _{0} ^{T} \rho \left( \phi \right) d \phi = 1$, and $\alpha = \dfrac{D\sigma _{\phi} ^2}{\Delta g}$ is the ratio of noise intensity to the coupling strength. The constant $A$ can be determined by the periodicity condition of $\rho _0$, that is, $\rho _0 \left( 0 \right) = \rho _0 \left( T \right)$. Therefore the final form of stationary solution is
\begin{eqnarray}
\rho \left( \phi \right) = \dfrac{1}{N} e^{M\left( \phi \right)}   \times   \left[ \frac{e^{-\frac{1}{\alpha} \left( \frac{T \Delta\omega}{\Delta g}  -1 \right)} - 1}{\int _{0}^{T} e^{-M\left( \tilde{\phi} \right)} d\tilde{\phi}} \int _{0}^{\phi } e^{-M \left( \tilde{\phi} \right) } d\tilde{\phi} + 1\right].
\label{Ststate}
\end{eqnarray}

In Figure 2A we have shown the result of analytic solution for steady-stat phase difference distributions \ref{Ststate} and that of direct numerical integration of of phase differential equations \ref{ophase}. 
For solving Eq. \ref{Ststate}, we have used double “int” function of MATLAB. In simulation, we integrate Eqs. \ref{ophase} with Euler method and save spike times of each neuron. Then we have used “hist” function in MATLAB to plot $\rho$. 

It has been shown that in the weak coupling and weak noise limit, the cross-crologram (CC) and the phase difference probability distribution function, $\rho (\phi)$, are related by
\begin{eqnarray}
C(\tau ) = \rho \left( \dfrac{\tau}{T} \right). 
\end{eqnarray}

The most probable phase difference of spiking of the neurons (location of the peak of the PDF in Fig. 2A) can be determined by differentiation of $\rho$ with respect to $\phi$. Derivative of $\rho$ with respect to $\phi$ is 
\begin{eqnarray}
\dfrac{d\rho}{d\phi} = \dfrac{dM(\phi )}{d\phi} e^{M(\phi )}\left[ \dfrac{e^{-\alpha ^{-1} \left( \frac{T\Delta\omega}{\Delta g} -1 \right)}-1}{\int _{0}^{T} e^{-M( \bar{\phi} )} d\bar{\phi }} \int _{0}^{\phi} e^{-M (\bar{\phi} )} d\bar{\phi} +1   \right] + e^{M( \phi )} \left[    \dfrac{e^{-\alpha ^{-1} \left( \frac{T\Delta\omega}{\Delta g} -1 \right)}-1}{\int _{0}^{T} e^{-M( \bar{\phi} )} d\bar{\phi }}   e^{-M( \bar{\phi} )}    \right].
\label{S36}
\end{eqnarray}
For $Z(\phi ) = 1- \cos (\phi )$, $M(\phi )$ would be
\begin{eqnarray}
M( \phi ) = \dfrac{1}{\alpha} \left( \dfrac{\Delta \omega }{\Delta g}\phi -\dfrac{1}{T} \phi +\dfrac{1}{T}\sin (\phi ) \right),
\end{eqnarray} 
and Eq. \ref{S36} reduces to 
\begin{eqnarray}
\dfrac{d\rho}{d\phi} = \dfrac{1}{\alpha}\left( \dfrac{\Delta\omega}{\Delta g} - \dfrac{1}{T} + \dfrac{1}{T} \cos (\phi ) \right) e^{M(\phi )}\left[ \dfrac{e^{-\alpha ^{-1} \left( \frac{T\Delta\omega}{\Delta g} -1 \right)}-1}{\int _{0}^{T} e^{-M( \bar{\phi} )} d\bar{\phi }} \int _{0}^{\phi} e^{-M (\bar{\phi} )} d\bar{\phi} +1   \right] + \dfrac{e^{-\alpha ^{-1} \left( \frac{T\Delta\omega}{\Delta g} -1 \right)}-1}{\int _{0}^{T} e^{-M( \bar{\phi} )} d\bar{\phi }},  
\label{S37}
\end{eqnarray}
therefore 
\begin{eqnarray}
0 = \dfrac{1}{\alpha}\left( \dfrac{\Delta\omega}{\Delta g} - \dfrac{1}{T} + \dfrac{1}{T} \cos (\phi ^{*} ) \right) e^{M(\phi ^{*} )}\left[ \dfrac{e^{-\alpha ^{-1} \left( \frac{T\Delta\omega}{\Delta g} -1 \right)}-1}{\int _{0}^{T} e^{-M( \bar{\phi} )} d\bar{\phi }} \int _{0}^{\phi ^{*}} e^{-M (\bar{\phi} )} d\bar{\phi} +1   \right] + \dfrac{e^{-\alpha ^{-1} \left( \frac{T\Delta\omega}{\Delta g} -1 \right)}-1}{\int _{0}^{T} e^{-M( \bar{\phi} )} d\bar{\phi }},  
\label{S38}
\end{eqnarray}
and 
\begin{eqnarray}
e^{M(\phi ^{*} )}\left[ \dfrac{e^{-\alpha ^{-1} \left( \frac{T\Delta\omega}{\Delta g} -1 \right)}-1}{\int _{0}^{T} e^{-M( \bar{\phi} )} d\bar{\phi }} \int _{0}^{\phi ^{*}} e^{-M (\bar{\phi} )} d\bar{\phi} +1   \right] = - \dfrac{e^{-\alpha ^{-1} \left( \frac{T\Delta\omega}{\Delta g} -1 \right)}-1}{\dfrac{1}{\alpha}\left( \dfrac{\Delta\omega}{\Delta g} - \dfrac{1}{T} + \dfrac{1}{T} \cos (\phi ^{*} ) \right)\int _{0}^{T} e^{-M( \bar{\phi} )} d\bar{\phi }}.
\label{S40}
\end{eqnarray} 
By using equations \ref{S40} and \ref{Ststate} we have
\begin{eqnarray}
\rho ( \phi ^{*}) =- \dfrac{1}{N}  \dfrac{e^{-\alpha ^{-1} \left( \frac{T\Delta\omega}{\Delta g} -1 \right)}-1}{\dfrac{1}{\alpha}\left( \dfrac{\Delta\omega}{\Delta g} - \dfrac{1}{T} + \dfrac{1}{T} \cos (\phi ^{*} ) \right)\int _{0}^{T} e^{-M( \bar{\phi} )} d\bar{\phi }},
\label{S41}
\end{eqnarray}
then $\phi ^{*}$ is
\begin{eqnarray}
\phi ^ {*} (\Delta \omega) &=& \langle \phi \rangle \nonumber \\ &=& \int _{0} ^{T} \phi ~\rho _{\Delta\omega} (\phi )d\phi.
\label{S42}
\end{eqnarray}
We have used MATLAB function "int" to plot $\phi ^{*}$ versus $\Delta\omega$ in Figure 2C using Eqs. \ref{Ststate}, \ref{S42}, and $C_{max}$; and $\rho _{max}$ versus $\Delta\omega$ in Figure 2B by Eq. \ref{S41}.

In Figure 3 we have plotted $C( \tau )$ for two LIF neurons described by
\begin{eqnarray}
\tau _{m}\dot{v_i} = v_{rest}-v_i +I_i + g_{ij}G(v_i,v_j)+\sqrt{D} \xi _i (t)
\label{LIFEQ}
\end{eqnarray}
with a reset condition ${v_i}(t^+ =V_{r})$ if ${v_i}(t^- >V_{th})$. we integrated this equation for two pulse coupled neurons and calculated spike count for a time window of $0.5 mS$. Parameters are selected in agreement with biological cases as $\tau_m=20~ms$, $v_{res}=-74~mV$, $V_{th}=-54~mV$, $v_r=-60~mV$, $I=25.0 +\Delta I~mV$, $g_{ij}=1~mV$ and $D=1.0$.


\end{document}